%% file: lattice06_05.tex
\documentclass{PoS}

\renewcommand{\vec}[1]{\vect{\rightarrow}{#1}{\:}}
\newcommand{\vect}[3]{{\mathchoice
    {\vecto{\displaystyle}{\scriptstyle}{#1}{#2}{#3}}%
    {\vecto{\textstyle}{\scriptstyle}{#1}{#2}{#3}}%
    {\vecto{\scriptstyle}{\scriptscriptstyle}{#1}{#2}{#3}}%
    {\vecto{\scriptscriptstyle}{\scriptscriptstyle}{#1}{#2}{#3}}}}
\newcommand{\vecto}[5]{\!\stackrel{{}_{{}_{#5 {#2 #3}}}}{{#1 #4}}\!}
\newcommand{\vdot}{\!\cdot\!}

\title{Charmed spectroscopy from a nonperturbatively determined
       relativistic heavy quark action in full QCD}

\ShortTitle{Charmed spectroscopy from a NP determined RHQ in full
QCD}

\author{\speaker{Huey-Wen Lin} %\thanks{}
        for the RBC Collaboration
        \\
        Department of Physics, Columbia University, New York, NY 10027\\
        Thomas Jefferson National Accelerator Facility, Newport News, VA 23606\\
        E-mail: \email{hwlin@jlab.org}}

\abstract{We present a preliminary calculation of the charmed
meson spectrum using the 2+1 flavor domain wall fermion lattice
configurations currently being generated by the RBC and UKQCD
collaborations. The calculation is performed using the
3-parameter, relativistic heavy quark action with
nonperturbatively determined coefficients. We will also
demonstrate a step-scaling procedure for determining these
coefficients nonperturbatively using a series of quenched, gauge
field ensembles generated for three different lattice spacings.}

\FullConference{XXIV International Symposium on Lattice Field Theory\\
         July 23-28 2006\\
         Tucson Arizona, US}

\begin{document}

\section{Introduction}
\vspace{-0.3cm} Flavor physics and CP violation play an important
role in particle physics. Lattice QCD provides a first-principles
approach for probing these interesting physics topics starting
from the Standard Model. However, in the application to heavy
quark physics, $(ma) \ll 1$ is no longer true and $(ma)^n$ terms
become significant. Because $a$ must be made very small, direct
simulation of heavy quarks by brute force becomes too expensive.
Thus, an effective theory is needed to carry out the calculation.
There are multiple fermion actions being used in lattice
calculations; see reviews in Refs.~\cite{Kronfeld:2003sd,
Wingate:2004xa,Okamoto:2005zg}.

In this work, we will concentrate on the effective theory called
relativistic heavy quark (RHQ) action
\cite{El-Khadra:1997mp,Aoki:2001ra,Christ:2006us,Christ:lat06}, as
\vspace{-0.2cm}
\begin{eqnarray}\label{eq:RHQ}
S & = & \sum_n \overline{\psi}_n \left\{  {m_0} + \gamma_0 D_0  -
\frac{1}{2} a D_0^2 + {\zeta}\right[ \vec{\gamma} \vdot \vec{D}
 - \frac{1}{2} a
\big(\vec{D}\big)^2\left] - \sum_i \frac{i}{4}{c_{\rm P}}
a\sigma_{\mu\nu}F_{\mu\nu} \right\}\psi_{n}.
\end{eqnarray}
We specifically use the formulation proposed in
Refs.~\cite{Christ:2006us,Christ:lat06}. The main idea is that in
the heavy quark case, the temporal covariant derivative $D_0$ is
around the order of $ma$ and should not be treated the same way as
the spatial derivatives $D_i$; we refer to this method as ``RHQ''
power counting. Following the Symanzik improvement procedure, we
found that there are only three necessary parameters in the
action: $m_0$, $\zeta$, $c_{\rm P}$. The advantages of using this
action are that it \vspace{-0.2cm}
\begin{itemize}
\item{Systematically absorbs mass factors into the coefficients}
\vspace{-0.2cm}
\item{Has small cutoff effects: $(\Lambda_{\rm QCD}a)^2$ for
    heavy-light systems and $(\alpha_s m a)^2$ for onium systems}
\vspace{-0.2cm}
\item{Goes to Sheikoleslami-Wohelert (SW) action when $ma\ll 1$ and
Non-Relativistic QCD %(NRQCD)
 when $ma\gg 1$}
%nhc: \item{Has no renormalon shadow\cite{Kronfeld:2003sd}}
\end{itemize}
\vspace{-0.2cm} In high-precision calculations, we must first
determine the correct action parameters before using the action to
calculate observables. Since perturbation theory introduces error
into our parameter calculation, a nonpertubatively determined
action is important.

\vspace{-0.2cm}
\section{Nonpertubatively determined RHQ action}
\begin{figure}[hbt]
\vspace{-0.5cm}
\includegraphics[width=0.9\columnwidth]{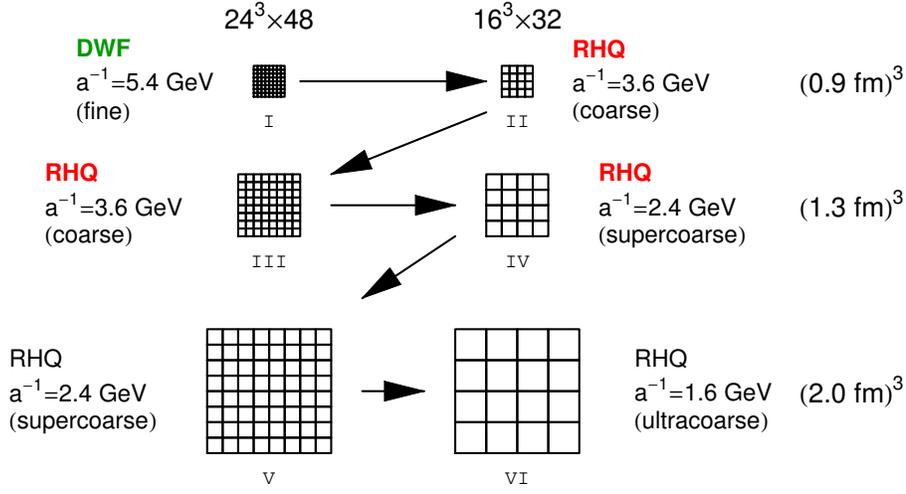}
\vspace{-0.3cm} \caption{The sequence of lattice sizes and lattice
spacings used to determine the coarse-lattice heavy quark
parameters through a step-scaling technique beginning with a
comparison with an $O(a)$-improved light quark calculation: The
matchings between the top four lattice spacing-volume combinations
are described in this work.}
\vspace{-0.3cm}\label{fig:step_scaling} \vskip -3mm
\end{figure}

\vspace{-0.4cm}
\label{sec:NPRHQ}%
There are various ways of determining the action coefficients. One
approach that has been widely used is lattice perturbation theory.
However, unlike continuum perturbation theory, LPT does not
converge quickly, and it has errors that are hard to control.
Another approach is to tune action parameters by matching %lattice
physical observables sensitive to particular parameters to their
known experimental values. However, we would lose some of the
predictive power of our theory. Last but not least, we may use a
step-scaling technique\cite{Luscher:1996jn}; instead of using
brute force to directly simulate the heavy quark on a fine lattice
(where $ma \ll 1$) in a large box, we connect fine lattices with small physical volumes to coarser
lattices with larger physical volumes, as shown in the
Figure~\ref{fig:step_scaling}. This is computationally much
cheaper than the brute-force calculation.

We can apply the step-scaling technique to determine the
coefficients in the action by using
\begin{enumerate}
\vspace{-0.2cm}
\item off-shell quantities: The quark-gluon vertex is a
candidate for determining the coefficients coupling to the
Sheikholeslami and Wohlert term. We will be able to use the
off-shell improvement to determine not only the coefficients in
the action, but also the $O(a)$ off-shell improvement coefficients
in the quark field, which allows us to apply nonperturbative
renormalization (NPR) directly\cite{Martinelli:1994ty,Lin:2005zd}.
However, this requires many more parameters to be determined
numerically and involves subtle, gauge--non-invariant terms to be
added to the action and operators. \vspace{-0.2cm}
\item on-shell quantities: We use mass combinations of
pseudoscalar~(PS), vector~(V), scalar~(S) and axial-vector~(AV)
mesons in heavy-heavy ($hh$) and heavy-light ($hl$) systems
computed in fixed, finite volume\cite{Lin:2006ur}.
\begin{enumerate}
\vspace{-0.2cm}
\item spin-averaged:  $m^{hh}_{\rm sa}=\frac{1}{4}\left(m^{hh}_{\rm PS} + 3
m^{hh}_{\rm V}\right)$,
$m^{hl}_{\rm sa}=\frac{1}{4}\left(m^{hl}_{\rm PS}+ 3 m^{hl}_{\rm V}\right)$%
\vspace{-0.1cm}
\item hyperfine splitting: $m^{hh}_{\rm hs}=m^{hh}_{\rm V} - m^{hh}_{\rm PS}
$, $m^{hl}_{\rm hs}= m^{hl}_{\rm V}-m^{hl}_{\rm PS}$%
\vspace{-0.1cm}
\item spin-orbit averaged and splitting: $m^{hh}_{\rm soa}
=\frac{1}{4}\left(m^{hh}_{\rm S} + 3 m^{hh}_{\rm AV}\right)$,
%\hspace{ 12cm}
\vspace{-0.1cm}
$m^{hh}_{\rm sos}=m^{hh}_{\rm AV} - m^{hh}_{\rm S} $%
\item dispersion relation: $E^2 = m^2 + c^2 p^2$.
%where $m_1$ is the rest mass and $m_2$ is kinetic mass.%
\end{enumerate}\vspace{-0.2cm}
\end{enumerate}\vspace{-0.2cm}

The simulation was carried out on a QCDOC 512-node machine at
420~MHz clock frequency. We used quenched Wilson gauge action with
the heatbath algorithm, taking 20,000 sweeps for thermalization
and measuring 100 configurations at 10,000 sweeps separation. 
Coulomb gauge-fixed hydrogenic source smearing is used to improve
overlap with ground states. 
We use a linear ansatz relating the action coefficients ($X_{\rm
RHQ}$) and the corresponding measurements ($Y_{\rm
coarse}^{i,d}$):
\begin{eqnarray}
Y_{\rm coarse}^{i,d} \;= \;A^d \;+ \;J^d \vdot X_{\rm RHQ}^i,%
\end{eqnarray}
where $J$ and $A$ can be obtained from either fitting parameters
or using finite differences directly from a Cartesian set.
These two approaches agree within errors. 
The detailed choices of parameters, lattice spacing determinations
and analysis can be found in Ref.~\cite{Lin:2006ur,Lin:thesis}.
The RHQ coefficients for the $a^{-1}=2.4$~GeV lattice, after two
stages of matchings starting
from the 
$a^{-1}=5.4$~GeV lattice, are
\begin{eqnarray*}\label{eq:XoutB6.074}
c_{\rm P} (m_0)  & =&  1.65(3)+ 0.12(6) m_0 + 1.06(4) m_0^2\\
\zeta (m_0)& =&  1.090(10) +0.318(16) m_0 -0.092(10) m_0^2
%,\\ m_0(a m_h)  & =& -0.536(24) +  3.80(13) a m_h+ 1.32(22) (a m_h)^2
.
\end{eqnarray*}

\begin{figure}
\begin{tabular}{cc}
\begin{minipage}{0.48\textwidth}
\vspace{-1.2cm}
\includegraphics[width=\columnwidth]{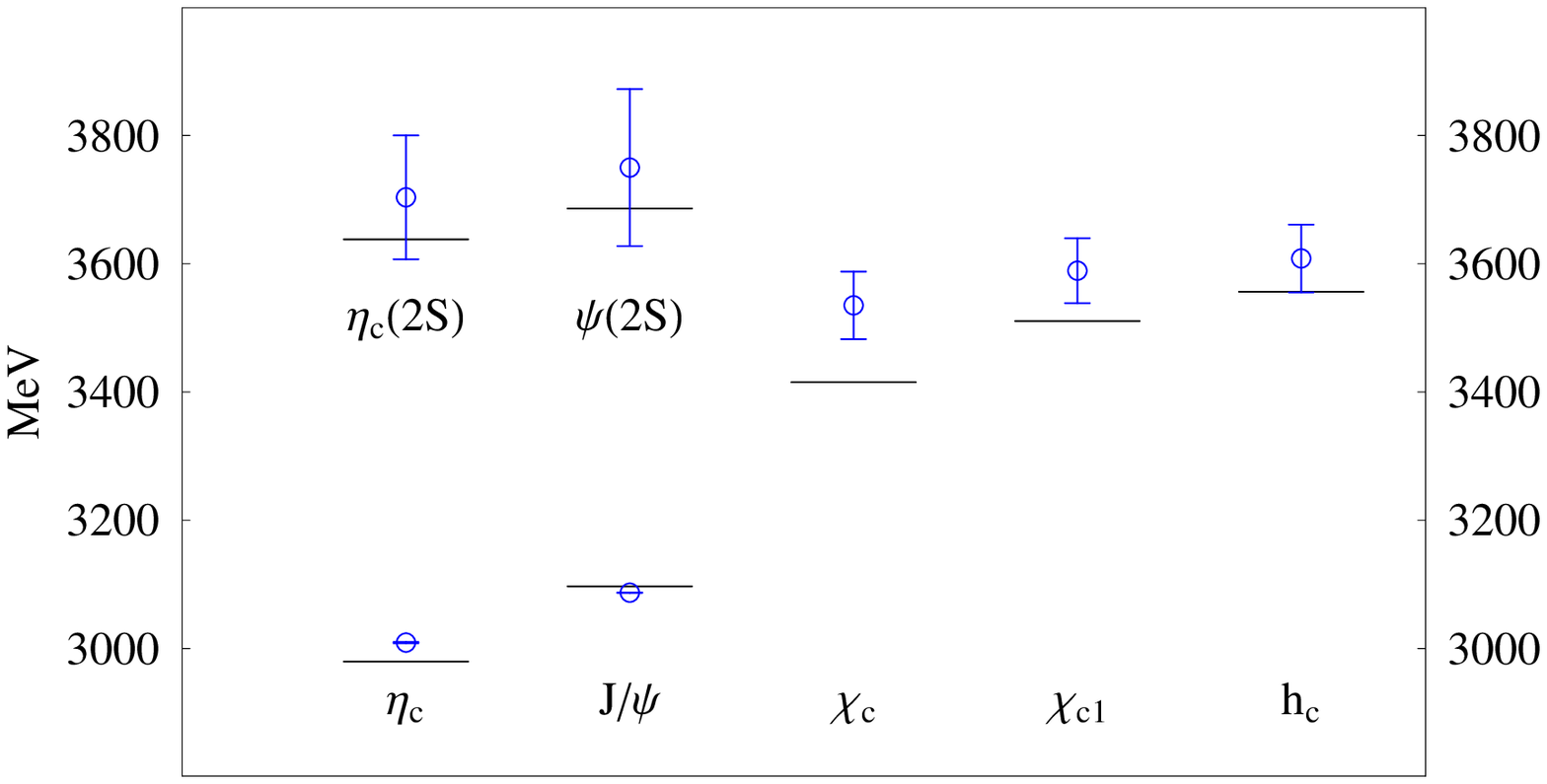}
\vspace{-1cm} \caption{The spectrum of the charmonium system: The
circles are our results with statistical errorbars and the
horizontal lines correspond to experimental values.} \vspace{-1
cm} \label{fig:charmSpec}
\end{minipage}
&
\begin{minipage}{0.48\textwidth}
\vspace{-0.8cm}
\includegraphics[width=\columnwidth]{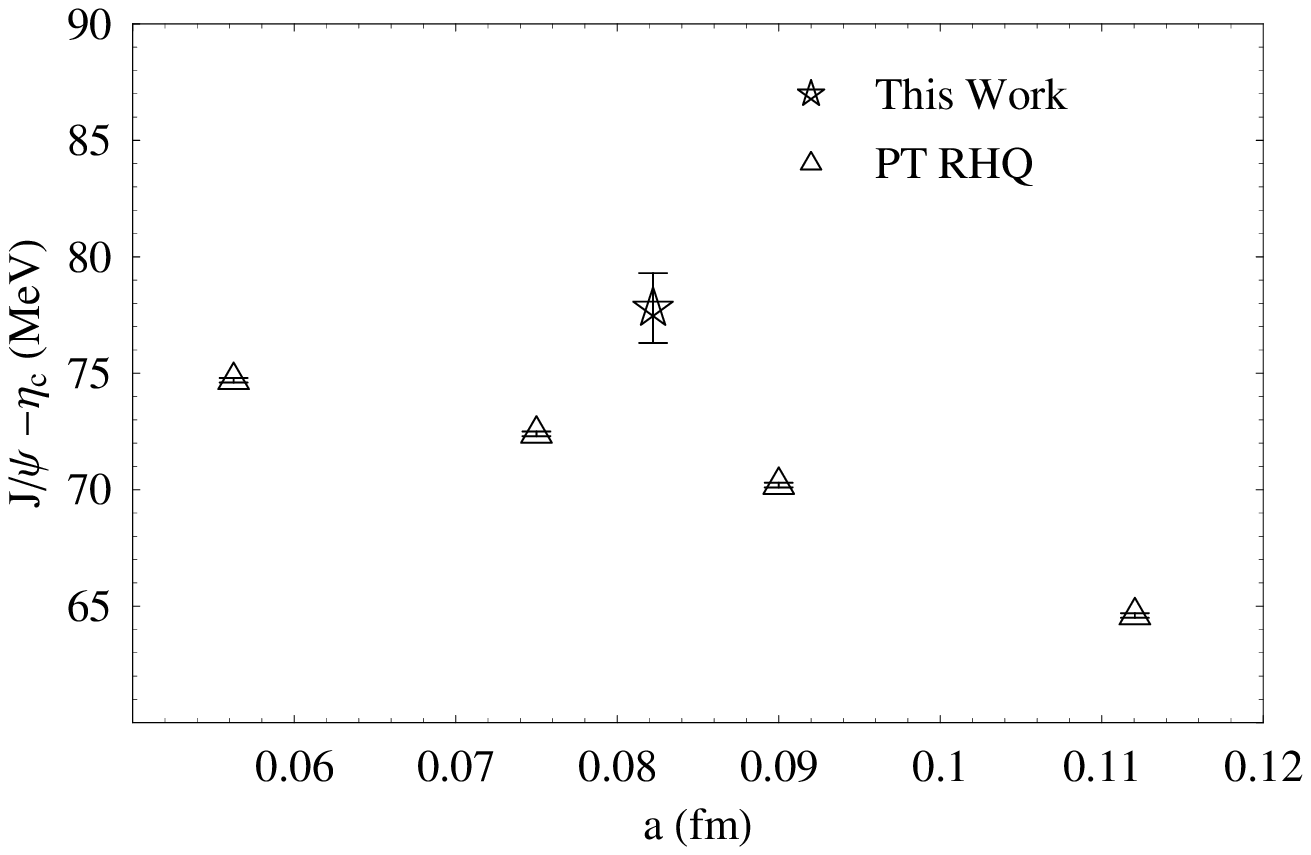}
\vspace{-0.8cm} \caption{A summary of hyperfine-splitting
calculations with quenched approximation using RHQ action.}
\label{fig:quenchedHypList} \vspace{-0.3cm}
\end{minipage}
\end{tabular}
\end{figure}

We simulate at three charm quark points from the RHQ coefficients,
$(X^{(3)}_{\cal C})^T = \{m_0, c_{\rm P}, \zeta \}=$ \{$-$0.06106,
1.651, 1.070\}, \{0.02173, 1.653, 1.097\}, \{0.1086, 1.686,
1.122\} respectively. The spin-averaged mass
($m_{\overline{1S}}=[m_{\eta_c}+3 m_{J/\Psi}]/4$) is used to
determine the bare charm quark mass. 
Figure~\ref{fig:charmSpec} shows a summary of our results along
with the experimental values, with statistical errors only. All
$P$-wave masses are within one standard deviation higher than the
experimental one. The spin-orbit splitting is 54(33)~MeV, similar
to what we observed in the past; it is a difficult quantity to
determine precisely. The hyperfine splitting is 77.8(15)~MeV,
about 40\% smaller than the experimental splitting (116~MeV). The
hyperfine splitting is typically not given correctly in lattice
calculations. A list of quenched hyperfine splittings calculated
from RHQ lattice QCD is given in Figure~\ref{fig:quenchedHypList}. The hyperfine splitting from the
one-loop Tsukuba approach\cite{Kuramashi:2005ww} ranges from 75.7
to 64.6~MeV using a fixed volume of $(1.8\mbox{ fm})^3$ with
lattice spacing varying from 0.0562 to 0.112. 
Our hyperfine splitting at the same lattice spacing is about 10\%
higher than the one-loop quenched result\cite{Kuramashi:2005ww},
which is encouraging. We might be able to resolve the hyperfine
splitting problem when we apply our nonperturbative approach in a
full-QCD calculation.

For the excited states, we use an additional smearing function on
the heavy quark field, \vspace{-0.2cm}
\begin{eqnarray}
\Psi_{\rm exc}(r) &=& (1 - r/2r_0)e^{-r/2r_0} \vspace{-0.3cm}
\label{eq:SmearFunc2},
\end{eqnarray}
to improve the overlap with excited states. Then, we perform a
four-parameter double-cosh fit to extract the ground and excited
states. The radial excited states from our low-statistics data
appear consistent with experiment, as does the
$\overline{2S}-\overline{1S}$ splitting. This measurement can be
easily improved with more configurations and a constrained fit.

So far in this discussion, we have been taking the lattice spacing
from the static quark potential with scale determined by Sommer
scale $r_0=0.5\mbox{ fm}$ from phenomenological models. We could
also determine the lattice spacing using the
$\overline{1P}-\overline{1S}$ splitting in the charmonium system.
Here we adopt the singlet $1P$ state, $m_{h_c}$, for the
$\overline{1P}$ mass. Recalculating the bare charm quark mass and
the lattice spacing, we get $am_{c}=0.15599(14)$ and
$a^{-1}=2.24(14)$~GeV. This suggests $r_0=0.534(33)$~fm. This is
close to what we assumed in the previous scale determination. From
the statistics we have, we do not observe problems due to
determining the lattice spacing in this calculation. Thus, we will
continue use $a^{-1}=2.4$~GeV for the rest of this work.

Fortunately, because of our step-scaling procedure, we are able to
calculate $Z_m^{\rm lat}$ in RI/MOM scheme\cite{Martinelli:1994ty}
using DWF on the fine lattice and to include an estimate of
finite-mass effects ($\sim (am_h)^2$) in the systematic error for
the final calculation.
We calculate the quark propagator and we expect that in the large-momentum region the scalar part of
the quark propagator behaves like\cite{Blum:2001sr}\vspace{-0.2cm}
\begin{eqnarray}\label{eq:FitForm}
 \frac{1}{12}{\rm Tr}(S_L^{-1}(p)) = C \times p^2 + Z_mZ_q(m +m_{\rm
res})
 + \frac{B}{p^2} + O(p^{-4}).
\end{eqnarray}\vspace{-0.cm}
When this equation is evaluated, we find the renormalized mass
$m_c^{\rm RI}=1.124(9)$~GeV in RI/MOM scheme.
We matched RI/MOM scheme to $\overline{\rm MS}$ scheme and
obtained the renormalization-group--invariant
mass\cite{Chetyrkin:1999pq}, $m^{\rm RGI}=2.401(21)$~GeV. We find
at the scale of averaged $m_c$, our renormalized charm quark mass
is $1.314(11)$~GeV.
%nhc
This statistical error is determined by applying the jackknife
method to the supercoarse statistical ensemble and then inflating
the result by a factor of $\sqrt{3}$ to include the statistical
errors introduced at the other two matching stages; thus,
$m^{\overline{\rm MS}}_c(\overline{m}_c)=1.314(18)$~GeV.

\vspace{-0.2cm}
\section{Full QCD}
\label{sec:fullQCD} \vspace{-0.3cm}

\begin{figure}
\begin{tabular}{cc}
\begin{minipage}{0.45\textwidth}
\vspace{-0.6cm}
\includegraphics[width=\columnwidth]{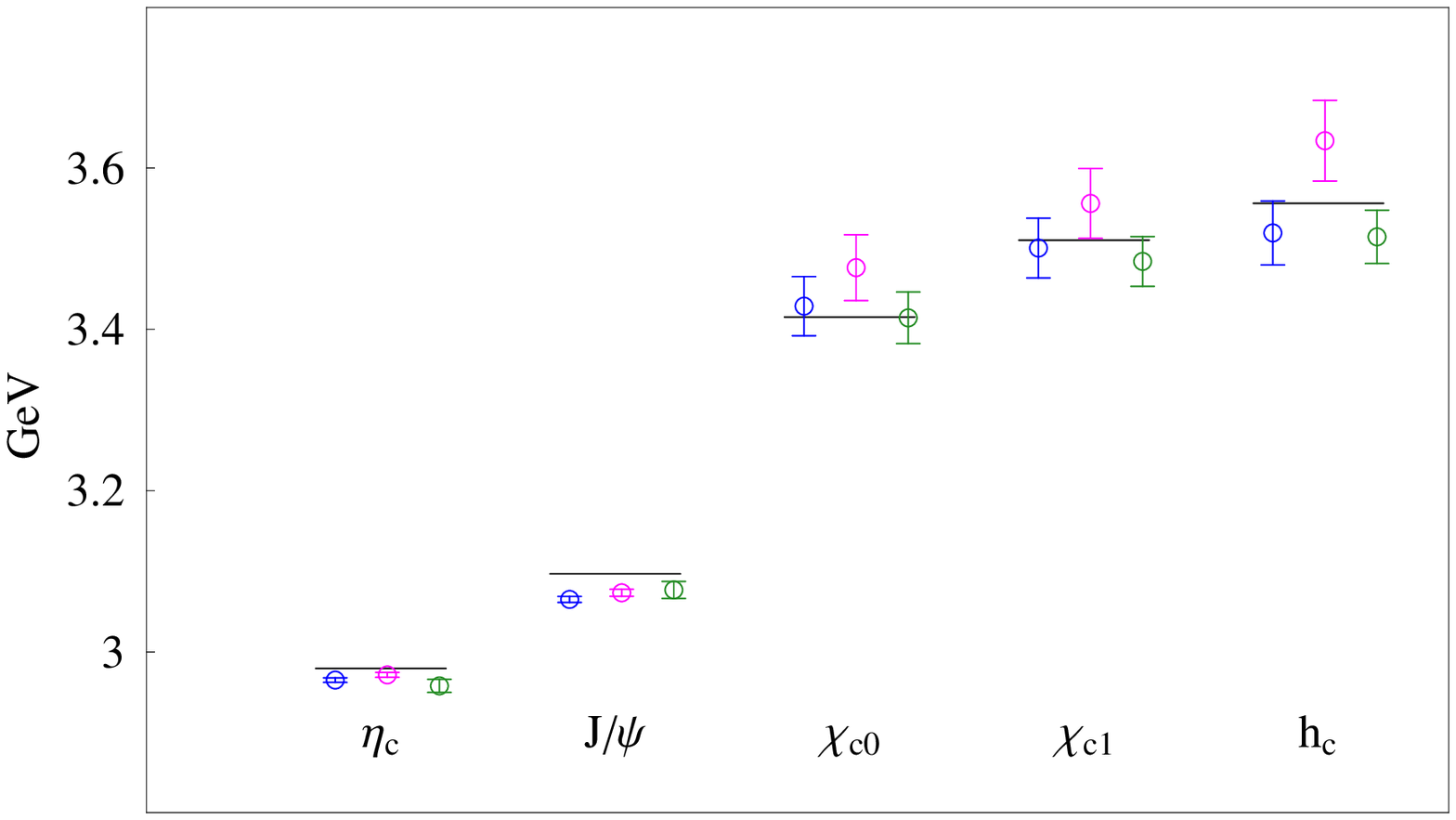}
\vspace{-0.8cm} \caption{The spectrum of the charmonium system
from NP RHQ in full QCD: The blue, purple and green correspond to
lattices with $am_{\rm sea}=0.01, 0.02, 0.03$ respectively, and
the horizontal lines correspond to experimental values.}
\label{fig:DynCharmSpec}
\end{minipage}
&
\begin{minipage}{0.5\textwidth}
\vspace{-1.2cm}
\includegraphics[width=\columnwidth]{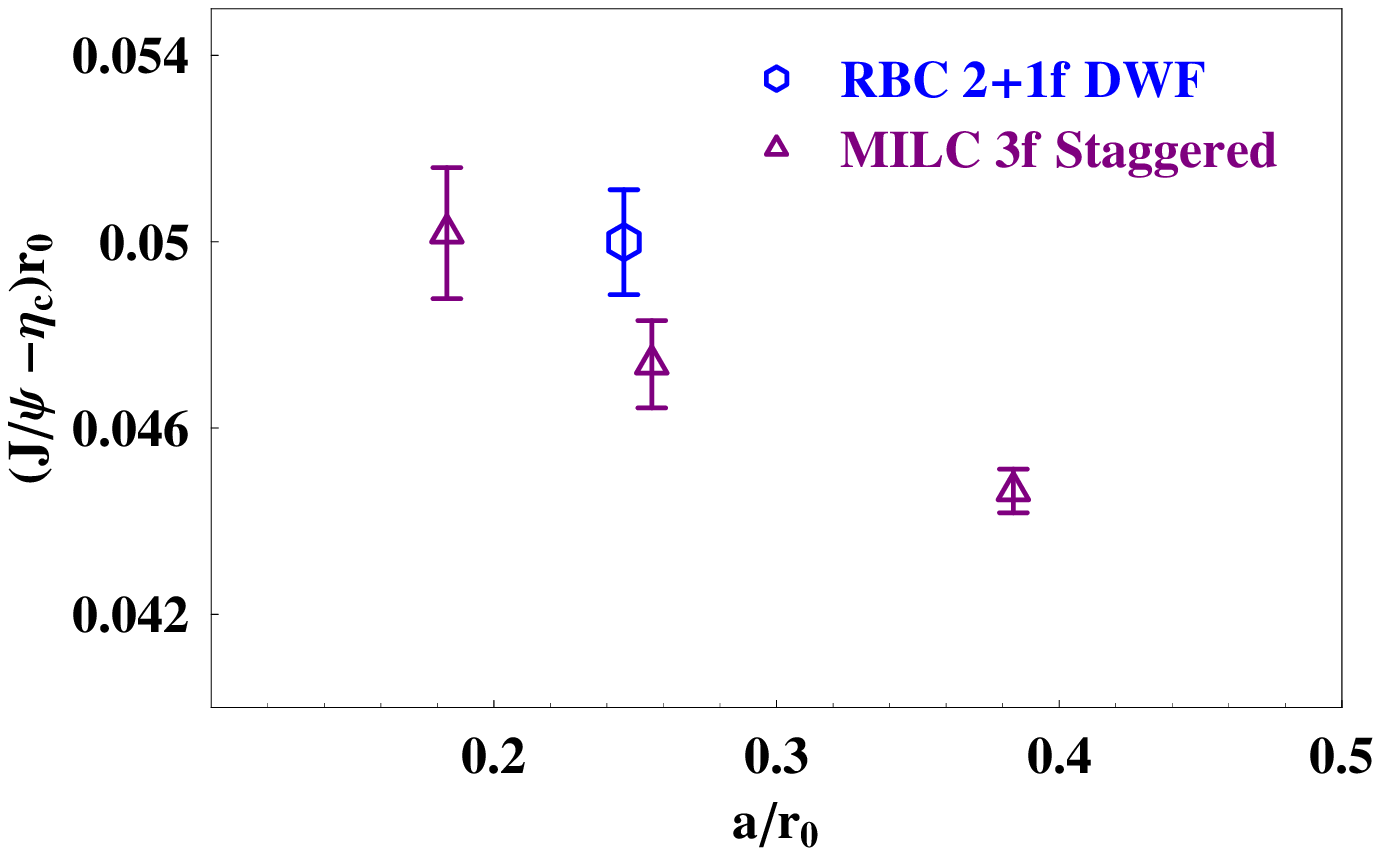}
\vspace{-0.5cm}  \caption{A comparison of hyperfine-splitting from
our $am_{\rm sea}=0.01$ and from MILC 2+1 dynamical calculations.
} \label{fig:DynHypList}\vspace{-0.3cm}
\end{minipage}
\end{tabular}
\end{figure}

The RBC and UKQCD collaborations have generated 2+1 flavor
dynamical DWF ensembles, at fixed lattice spacing $a^{-1} \approx
1.6$~GeV (set by Sommer scale 0.5~fm), with two volumes, $\approx
2^3$ and $3^3$~${\rm fm^3}$, using Iwasaki gauge action
($\beta=2.13$). The up and down sea quark mass in terms of the
inverse lattice spacing are 0.01, 0.02 and 0.03 with the strange
quark mass set to 0.04.
We first apply the NP RHQ coefficients %from quenched approximation
directly on 2+1-flavor dynamical configurations to check out the
heavy-heavy sector of charm physics, where the small volume
lattice is sufficient. The results are shown in
Figure~\ref{fig:DynCharmSpec} for three different sea quark
masses, labeled by blue, purple and green as the mass increases,
with statistics of 75 configurations. First, we can see that there
is an instant boost in the value of the hyperfine splitting
between quenched lattices and dynamical ones, and the $P$-wave
states all agree with the experiment values within statistics
error bars. In Figure~\ref{fig:DynHypList}, we compare our
hyperfine splitting with the calculation obtained from
MILC\cite{Gottlieb:2005me}. Since MILC and RBC/UKQCD use different
reference scales, Sommer scale $r_0$, to set the lattice spacing
and the hyperfine splitting is sensitive to this reference scale,
we plot the dimensionless quantity (the product of hyperfine
splitting and $r_0$) as a function of $a/r_0$. We noted that even
though our lattice is coarser than the MILC ones, our hyperfine
value is consistent with their finest point. This is a very
encouraging result to test on our NP coefficients. However, since
the gauge actions used in the quenched and dynamical simulations
are not matched, there might be $O(a^2)$ effects in the gauge
action. Thus, our hyperfine splitting can be further improved if
we carry out the NP RHQ project dynamically.

\begin{figure}
\begin{tabular}{cc}
\begin{minipage}{0.48\textwidth}
\vspace{-0.6cm}
\includegraphics[width=\columnwidth]{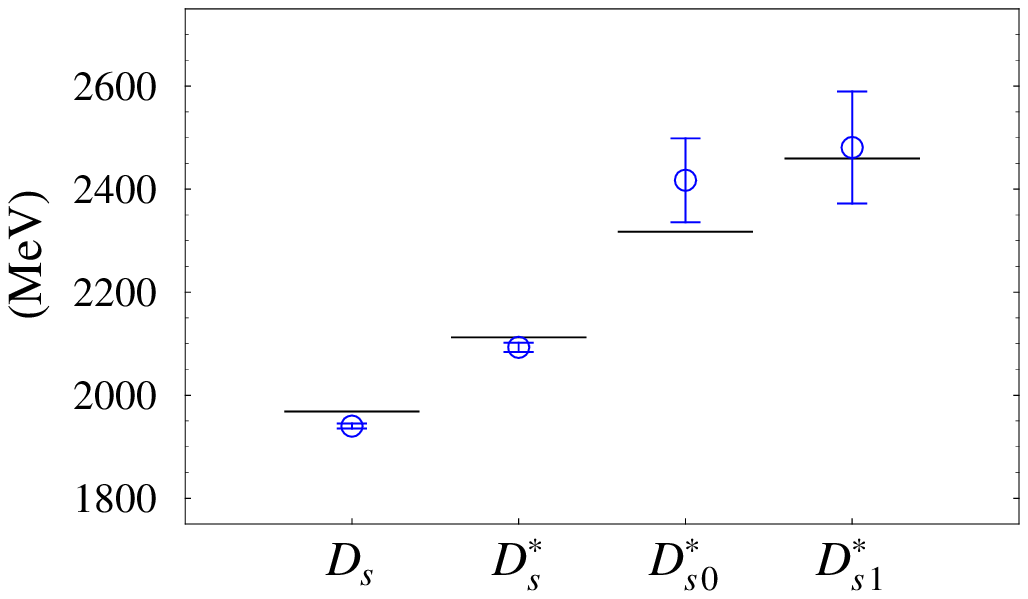}
\vspace{-0.5 cm} \caption{(Left) The spectrum of $D_s$ on the
$m_{\rm sea}=0.01$ ensemble with $m_{\rm strange}=0.04$.}
\label{fig:DynDsSpec}
\end{minipage}
&
\begin{minipage}{0.48\textwidth}
\vspace{-0.8cm}
\includegraphics[width=\columnwidth]{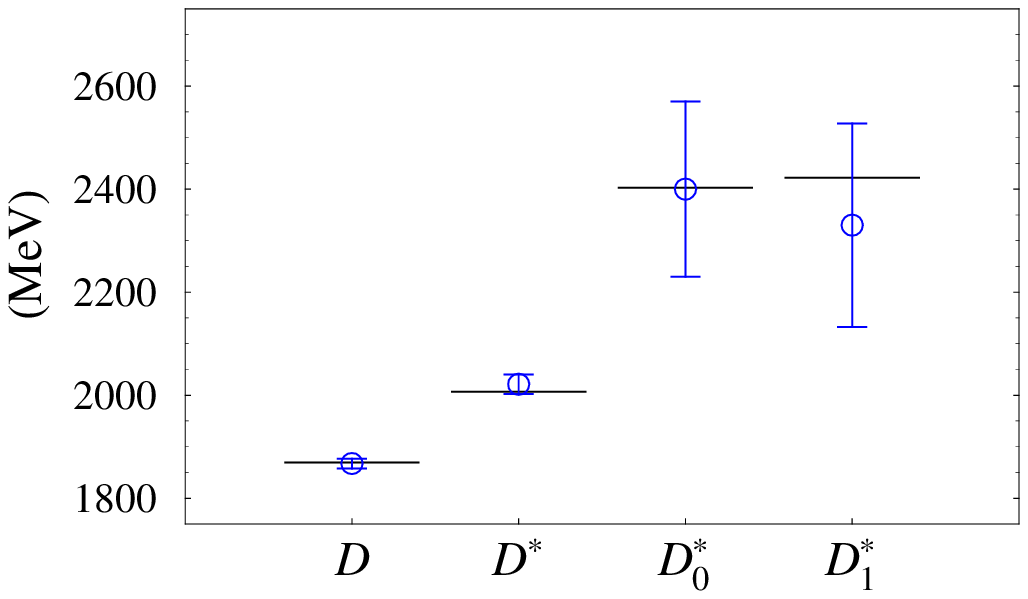}
 \vspace{-0.5cm} \caption{(Right) The
spectrum of $D$ on the $m_{\rm sea}=0.01$ ensemble with $m_{\rm
up,down}=0.01$. \vspace{-0.3cm}} \label{fig:DynDSpec}
\end{minipage}
\end{tabular}
\end{figure}

\begin{figure}
\begin{center}
\includegraphics[width=0.8\columnwidth]{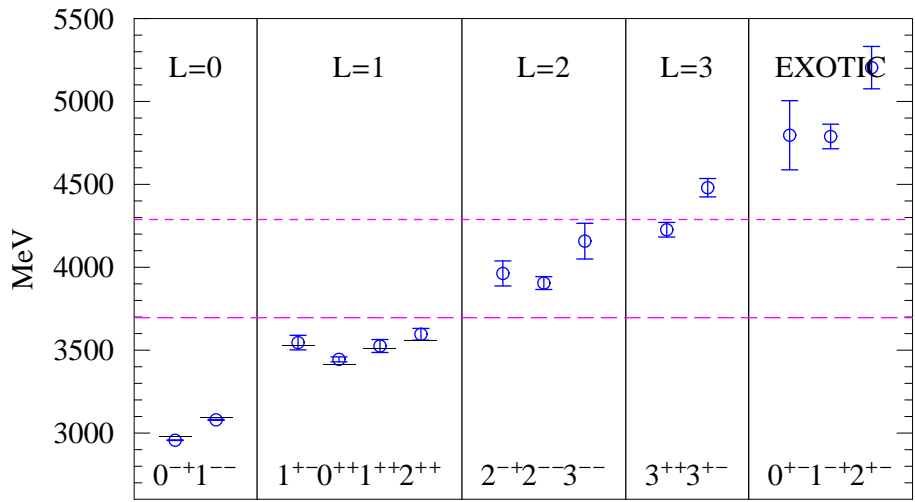}
\vspace{-0.5 cm} \caption{A preliminary study of the charmonium spectrum including operators with derivatives. The spectrum is classified by orbital angular momentum. The blue circles are our results and the black solid lines indicate the corresponding experimental values, if they exist. The two horizontal dashed lines indicate the $D^{**}D$ (top) and $D$ $\overline{\rm D}$ (bottom) thresholds.}
\label{fig:exotic}
\end{center}
\end{figure}

Carrying out a step-scaling calculation in full QCD is very
expensive and time-consuming; in the meantime, we can match the
RHQ coefficients on the RBC/UKQCD full QCD lattices to the
experimental charmonium spin-averaged mass and spin-splitting mass
and dispersion relation. This is accurate through $|\vec p a|$ and
to all order of $(ma)^n$. Therefore, we can immediately start
studying other charmed states, such as charmed baryons,
charm-light systems, charmonium excited states (including exotic
ones). We first demonstrate on our smallest sea quark mass
configurations; we obtained coefficients in the action
\vspace{-0.4cm}
\begin{eqnarray}
X_{\rm RHQ} = \{m_0,c_{\rm P},\zeta\}=\{  0.46(4), 2.50(9),
1.285(19)\}. \vspace{-0.5cm}
\end{eqnarray}
Again, the bare charm mass is set by the
experimental charmonium spin-averaged mass. We use 126
configurations. First, we look at the
results for the charm-light system, as shown in
Figures~\ref{fig:DynDsSpec} and \ref{fig:DynDSpec}. Both spectra
are consistent with experimental values. The $S$-wave splittings
are %\vspace{-0.5cm}
\begin{eqnarray*}
m_{D^*}     - m_D         = 154(16) \mbox{ MeV} \\
m_{D^*_{s}} - m_{D_{s}} = 153(7) \mbox{ MeV},
\end{eqnarray*}
%\vspace{-0.5cm}
consistent with the experimental values
of 142~MeV and 144~MeV respectively.
%
%\vspace{0.5cm}
Next, we look at higher excited states in the charmonium system, some of which involve non-local operators. The details of the operators used in this calculation can be found in Ref~\cite{Liao:2002rj}. Figure~\ref{fig:exotic} shows our preliminary results classified by orbital angular momentum. For $L=1$, our results are nicely consistent with the experimental values; we also make predictions for the higher-$L$ and gluonic exotic states. This might be helpful for identifying some mysterious experimental observations, such as $Y$(3940), $Y$(4260) and $X$(3872).

\vspace{-0.cm}
\section{Conclusion}
\label{sec:conclusion} \vspace{-0.3cm}
In this work, we have shown that the three coefficients ($m_0$,
$c_{\rm P}$ and $\zeta$) for the RHQ action can be
determined nonperturbatively with step scaling. %
We demonstrate some interesting quenched physics with improvement
of the hyperfine splitting in charmonium. It also allows us to
determine the charm quark mass: $m^{\overline{\rm
MS}}_c(\overline{m}_c)=1.314(18)$~GeV, without complicated
off-shell improvement on the RHQ action. This makes future
application to full QCD feasible. The direct application of NP
quenched coefficients to dynamical configurations gives promising
charmonium features such as better hyperfine splitting than
existing dynamical results. We further tune the RHQ action with
experimental numbers and produce promising results for the
charmonium and charm-light systems. In the near future, we would
also like to calculate more meson states involving derivative
operators (including exotic ones) and also to study charmed baryons masses. It should also be easy 
to apply this method to the $B$ system. In the long term, we would
like to carry on the step-scaling technique first to the NP
determined RHQ action in full QCD and then to decay constants and
form factors involving one or two heavy quarks with RI/MOM NPR.

\vspace{-0.4cm}
\section*{Acknowledgements}
\vspace{-0.4cm} We acknowledge helpful discussions with Tanmoy
Bhattacharya, Peter Boyle, Paul Mackenzie, Andreas Kronfeld, Sinya
Aoki, Yoshinobu Kuramashi and members of the RBC collaboration.
 We also thank Thomas Manke and Xiaodong Liao for their help with nonlocal operators in the charmonium system. %
In addition, we thank Peter Boyle, Dong Chen, Mike Clark, Norman
Christ, Calin Cristian, Zhihua Dong, Alan Gara, Andrew Jackson,
Balint Joo, Chulwoo Jung, Richard Kenway, Changhoan Kim, Ludmila
Levkova, Xiaodong Liao, Guofeng Liu, Robert Mawhinney, Shigemi
Ohta, Konstantin Petrov, Tilo Wettig and Azusa Yamaguchi for
developing the QCDOC machine and its software. This development
and the resulting computer equipment used in this calculation were
funded by the U.S. DOE grant DE-FG02-92ER40699, PPARC JIF grant
PPA/J/S/1998/00756 and by RIKEN.

This work was supported by DOE grant DE-FG02-92ER40699, PPARC
grant PP/C504386/1 and PPARC grant PP/D000238/1. We thank RIKEN,
BNL and the U.S. DOE for providing the facilities essential for
the completion of this work.

\vspace{-0.3cm}
\input{lattice06_05.bbl}
\vspace{-0.5cm}
\end{document}

%% file: lattice06_05.bbl
\providecommand{\href}[2] {#2}\begingroup\raggedright\endgroup